\documentclass[aps,prb,twocolumn]{revtex4}
\usepackage{graphicx,color}
\usepackage{amsmath}
\usepackage{amssymb,bm}

\renewcommand{\Re}{\mathop{\mathrm{Re}}}
\newcommand{\f}[1]{Fig.~\ref{#1}}
\newcommand{\eq}[1]{Eq.~(\ref{#1})}
\newcommand{\eqs}[2]{Eqs.~(\ref{#1}) and~(\ref{#2})}
\newcommand{\eqss}[3]{Eqs.~(\ref{#1}), (\ref{#2}) and~(\ref{#3})}

\def\d{\partial}
\def\be{\begin{equation}}
\def\ee{\end{equation}}
\def\bea{\begin{eqnarray}}
\def\eea{\end{eqnarray}}
\def\l({\left(}
\def\r){\right)}

\begin{document}

\bibliographystyle{apsrev}
\title{Finger patterns produced by thermomagnetic instability in
superconductors}

\author{A. L. Rakhmanov$^1$, D.~V. Shantsev$^{2,3}$, Y. M.
Galperin$^{2,3}$,
T.~H.~Johansen$^{2,4,}$\cite{0}}

\address{
$^1$ Institute for Theoretical and Applied Electrodynamics,
Izhorskaya 13/19, Moscow, 125412, Russia\\
$^2$ Department of Physics, University of Oslo, P. O. Box 1048
Blindern, 0316 Oslo, Norway\\
$^3$ A. F. Ioffe Physico-Technical Institute, Polytekhnicheskaya
26, St.Petersburg 194021, Russia\\
$^4$ Texas Center for Superconductivity and Advanced Materials,
University of Houston, Houston Texas, 77204-5002 USA}
\date{\today}

\begin{abstract}
A linear analysis of thermal diffusion and Maxwell equations is
applied to study the thermomagnetic instability in a type-II
superconducting slab. It is shown that the instability can lead to
formation of spatially nonuniform distributions of magnetic field
and temperature. The distributions acquire a finger structure with
fingers perpendicular to the screening current direction. We
derive the criterion for the instability, and estimate its
build-up time and characteristic finger width. The fingering
instability emerges when the background electric field is larger
than a threshold field, $E>E_c$, and the applied magnetic field
exceeds a value $H_{\text{fing}} \propto 1/\sqrt{E}$. Numerical
simulations support the analytical results, and allow to follow
the development of the fingering instability beyond the linear
regime. The fingering instability may be responsible for the
nucleation of dendritic flux patterns observed in superconducting
films using magneto-optical imaging.
\end{abstract}


\maketitle

\section{Introduction} \label{In}

The thermomagnetic instability or flux jumping is commonly
observed at low temperatures in type-II superconductors with
strong pinning.~\cite{MR,GMR,wipf91,chapter} The instability
arises because of two fundamental reasons: (i) motion of magnetic
flux releases energy, and hence increases the local temperature;
(ii)
the temperature rise decreases flux pinning, and hence facilitates
the flux motion. This positive feedback can result in thermal
runaways and global flux redistributions jeopardizing
superconducting devices. This mechanism was understood in early
works,\cite{swartz,Wipf} and later on the thermomagnetic
instability was studied thoroughly (see
Refs.~\onlinecite{MR,GMR,wipf91,chapter} for a review). In
particular, the threshold magnetic field for the instability was
calculated and its experimentally found dependence on temperature,
sample dimensions, and the applied field ramping rate were
explained.

The conventional theory of the thermomagnetic
instability\cite{MR,GMR}
predicts ``uniform'' flux jumps, where the flux front is
essentially flat. In
other words, the spatial extension of the instability region tends
to be maximal since small-scale perturbations are stabilized by
thermal diffusion. This picture is true for many experimental
conditions, however, not for all.
Numerous magneto-optical studies have revealed that the
thermomagnetic
instability in thin superconducting samples 
results in dendritic flux
patterns.\cite{1967,leiderer,bolz,bolz03,duran,vv,epl,sust,apl,prb,phc04,rudnev}
In course of the dendritic instability the flux forms narrow
``fingers'' of typical width 20--50~$\mu$m and length up to the
size of the sample. Such a behavior clearly contradicts to the
conventional theoretical concepts and needs elucidation.

Few attempts to describe a nonuniform development of
the thermomagnetic instability have been made. Among them is a
numerical solution of thermal diffusion and Maxwell equations that
can result in a rather nonuniform temperature distribution for a
bulk superconductor.~\cite{GV} Molecular dynamics simulations of
flux quanta motion in superconducting film\cite{epl} can model
dendritic flux and temperature patterns similar to those found
experimentally. However, these numerical results still lack
analytical support. In particular, it is still unclear what kind
of spatial structure can be formed during a flux jump, and under
what conditions. A similar problem was analyzed in a recent
work\cite{shapiro} where the propagating flux front was shown to
acquire a non-uniform spatial structure if its speed is higher
than some
critical value, and the conductivity is a strong function of flux
density. In the present study it is shown that these assumptions
are not necessary requirements for a superconductor to develop
nonuniform flux jumps.

In the present paper the spatial pattern of the instability in a
bulk superconductor is studied using  the conventional
approach\cite{MR,GMR,Wipf} -- linear analysis of a set of
differential equations describing small perturbations in the
electric field $E$ and temperature $T$. In contrast to the
previous investigations, we allow the perturbations to vary in any
direction, i. e., both parallel and perpendicular to the direction
of the background current $\mathbf{j}$ and field $\mathbf{E}$. In
this way we determine the stability criteria and also estimate the
instability build-up time. As a main result we find that the most
unstable perturbations are in the form of narrow fingers
perpendicular to the background field $\mathbf{E}$ and occur if
$E$ is larger than some threshold value. This shape prevents
current adjustment across the perturbed region and, hence, yields
the fastest perturbation growth. Too narrow fingers are, however,
suppressed by the thermal diffusion. Thus, the typical finger
size, $\sqrt{\kappa (\partial j_c/\partial T)^{-1}/E}$, where
$\kappa$ is the thermal conductivity and $j_c$ is the critical
current density, is determined by the competition between the
Joule heat $jE$ and thermal diffusion, $\kappa \nabla^2 T$.

\section{Basic equations} \label{Beq}

\begin{figure}[t]
\includegraphics [width=0.4\textwidth]{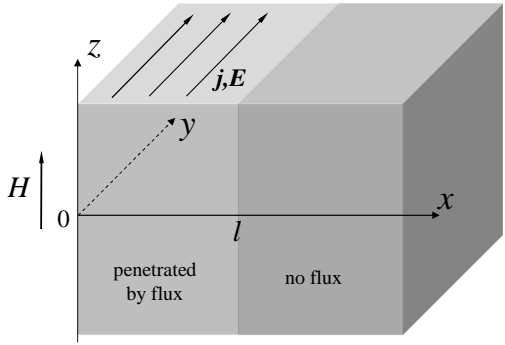}
\caption{\label{fig1} Superconductor geometry.}
\end{figure}

We shall study the instability in the simplest geometry, i.~e., in
a superconducting slab placed in a parallel magnetic field, see
Fig.~\ref{fig1}. The slab fills the semispace $x>0$, and the
external magnetic field $\mathbf{H}$ is parallel to the $z$-axis
so that the screening current $\mathbf{j}$ flows along the
$y$-axis. The current and magnetic field distributions in the
sample are determined by the Maxwell equation, with a proper
boundary condition
\begin{equation}
\label{B} \textrm{curl }\mathbf{B}=\mu_0 \mathbf{j},\,\,\
\mathbf{B}\vert_{x=0}
= \mu_0 \mathbf{H}.
\end{equation}
Here the local magnetic field in the flux penetrated part of the
slab is
assumed larger than
the first critical field, and hence, to a good approximation
$B(x,y)=\mu_0 H(x,y)$.
To find the temperature and electric field in a superconductor the
corresponding thermal and the second Maxwell equations should be
used:
\begin{eqnarray}
C(\partial T/\partial t)&=&\kappa \nabla^2
T+\mathbf{j}\mathbf{E}\, , \label{th} \\
\textrm{curl } \mathbf{E}&=& -\partial \mathbf{B}/\partial t\, ,
\label{E}
\end{eqnarray}
where $C$ is the specific heat.

These equations should be supplemented by a current-voltage curve
$j=j(E,B,T)$. In type-II superconductors the $j(E)$ dependence is
strongly nonlinear. As a result, a quasi-static critical state
with $j \approx j_c(B,T)$ is formed. This will be the initial
state from which the instability evolves. For simplicity we use
the
Bean model, i.~e. we neglect any $B$ dependence of the critical
current density $j_c$.
The exact form of the current-voltage curve,
\be
\mathbf{j} = j(T,E)\, (\mathbf{E}/E)\, .
\label{cvcf}
\ee
is not crucially important. The only important point is that the
$E(j)$ curve
is very steep, and therefore its logarithmic
derivative is large:
\begin{equation}\label{n}
n(E)  \equiv \frac{\d \ln E}{\d \ln j} \approx \frac{j_c}{\sigma
E} \gg 1 .
\end{equation}
Here $\sigma$ is the differential conductivity,
\begin{equation}\label{sigma}
\sigma(E)\equiv
\partial j / \partial E\, .
\end{equation}
At low electric fields, the $E(j)$ curve is often approximated by
a power law, i. e., $n$ is assumed
independent of $E$, and $E \propto j^n$. Our approach is
applicable also to the flux flow
regime at high electric fields. In that regime
$\sigma(E)=\sigma_{f}$ is the flux-flow Ohmic conductivity and
$n(E)=j_c/\sigma_{f}E \propto 1/E$.

The key dimensionless parameter of the model is the ratio of
thermal and magnetic diffusion coefficients:
\begin{equation}\label{tau}
\tau \equiv \mu_0 \sigma \kappa /C \, .
\end{equation}
The smaller $\tau$ is, the slower heat diffuses from the
perturbation region into the surrounding areas. Hence, one can
expect that for smaller $\tau$; (i) the superconductor is more
unstable, and (ii) the formation of instability-induced
nonuniform structures
is more favorable.


\section{Perturbation analysis} \label{Pan}

\subsection{Linearization of the problem}

We seek solutions of the equations presented above in the form,
\begin{equation}\label{per}
T+\delta T(x,y,t), \quad \mathbf{E}+\delta \mathbf{E}(x,y,t),
\end{equation}
where $T$ and $\mathbf{E}$ are the background values. The
background field $\mathbf{E}$ may be created by ramping the
external magnetic field $H$, or by other sources as discussed in
Sec.~\ref{Disc}. In practice $\mathbf{E}$ is nonuniform,
but for simplicity we disregard its coordinate dependence.
For a weak non-uniformity that can be justified using the method
of
Ref.~\onlinecite{mr76}, based on Wentzel-Kramers-Brillouin
approximation. In this approximation the non-uniformity results
only
in replacement some of local quantities by the ones averaged over
$x$. Hence, we get only insignificant numerical corrections.
Numerical simulations in Sec. V show that this conclusion also
holds in the
realistic situation when the non-uniformity of $\mathbf{E}$ is
induced
by the by ramping the external magnetic field $H$.
Similarly, we ignore any
coordinate dependence of the background temperature. This can be
done if it satisfies the inequality $T(x,y)-\overline{T}\ll
T_c-\overline{T}$, where $T_c$ is the critical temperature of the
superconductor, and $\overline{T}$ is the sample-averaged
temperature
before the instability build-up.

From the symmetry of the problem $E_x=0$, while for the
perturbation $\delta\mathbf{E}$ both $\delta E_x$ and $\delta E_y$
in general do not vanish. The linearization of the $E(j)$ in
(\ref{cvcf}) yields
\begin{equation}\label{CVCl}
\delta\mathbf{j}=\left(\frac{\partial j_c}{\partial T}\, \delta T
+
\sigma\, \delta E\right)
\frac{\mathbf{E}}{E}+j_c\left(\frac{\delta\mathbf{E}}{E}-\delta E
\frac{\mathbf{E}}{E^2}\right).
\end{equation}
Since the vector $\mathbf{E}$ is parallel to the $y$-axis,
one has
in the linear approximation that $\delta E =\delta E_y$, and as a
result
one finds
\begin{equation}\label{CVCL}
\delta\mathbf{j}=\left(\frac{\partial j_c}{\partial T}\, \delta T
+ \sigma \, \delta E_y \right)
\frac{\mathbf{E}}{E}+j_c\frac{\delta\mathbf{E}_x}{E}\, .
\end{equation}

We shall seek perturbations in the usual form:
\begin{eqnarray} \delta T &=&T^{*}\theta
\exp(\lambda t/t_0+ik_y\eta+ik_x\xi)\, , \label{per1}\\
\delta E_{x,y}&=&E\varepsilon_{x,y}\exp(\lambda
t/t_0+ik_y\eta+ik_x\xi)\, , \label{per2}
\end{eqnarray}
where $\xi=x/w$, $\eta=y/w$, and
\begin{equation}\label{variables}
\frac{1}{T^*}=-\frac{1}{j_c}\frac{\partial j_c}{\partial T}, \,\,
t_0 = \frac{\sigma CT^* }{j_c^2}=\mu_0 \sigma w^2, \,\,
w^2 = \frac{CT^*}{\mu_0j_c^2}\, . \nonumber
\end{equation}
Here $\theta$ and $\varepsilon_{x,y}$ are the Fourier amplitudes, 
$\Re\lambda$ is the dimensionless instability increment,
$t_0$ is the characteristic time of adiabatic
heating, which coincides with the magnetic diffusion time for the
length $w$, and $w$ is the characteristic scale for the adiabatic
instability.\cite{MR} The wave numbers $k_y$ and $k_x$
characterize the scale of the perturbation along the $y$ and $x$
axes, respectively. Since the sample is assumed infinite in the
$y$ direction,
the $k_y$ is arbitrary, while $k_x$ is determined by the width of
the flux penetrated region and the corresponding boundary
conditions.

Let us define the Fourier amplitude of the dimensionless current perturbation 
$\delta\mathbf{j}/j_c$ as $\mathbf{i}$. Using
Eqs.~(\ref{CVCL})--(\ref{per2}) one finds the components of the
vector $\mathbf{i}$ in the form
\begin{equation}\label{i}
 i_x=\varepsilon_x, \,\, i_y=-\theta+ \varepsilon_y/n
\, .
\end{equation}

Using Eq.~(\ref{th}) one obtains the equation for the temperature
perturbation $\theta$ as
\begin{equation}\label{thL}
\lambda\theta=-\tau (k_y^2+k_x^2)\theta+(i_y+\varepsilon_y)
/n \, .
\end{equation}
We find from Eq.~(\ref{thL})
\begin{equation}\label{T-E}
\theta=\frac{(1+1/n )\varepsilon_y}{n\lambda+n\tau
(k_y^2+k_x^2)+1}\, .
\end{equation}

Then, using
Eqs.~(\ref{B}) and (\ref{E}), we can rewrite the Maxwell equation
for the perturbation as
\begin{equation}\label{Eldyn}
{\bf k} \times[{\bf k} \times \vec{\varepsilon}\,]=\lambda n \,
 \mathbf{i}\, .
\end{equation}
Using the relations (\ref{i}) we cast Eq.~(\ref{Eldyn}) into the
equation set for dimensionless components of the electric field
perturbation
\begin{eqnarray}
&&\varepsilon_x=\frac{k_y k_x}{k_y^2+\lambda n} \, \varepsilon_y\,
, \label{Ex} \\
&& -k_x^2 \varepsilon_y + k_y k_x \varepsilon_x=\lambda
n(-\theta+\varepsilon_y/n)\, .  \label{Ey}
\end{eqnarray}
Note that these equations together with \eqs{i}{T-E} provide
continuity of the current perturbation, i. e., $ \textrm{div}\,
\mathbf{\delta j}=0$, as required. Substituting Eqs. (\ref{T-E}) and
(\ref{Ex}) in Eq.~(\ref{Ey}) one finds the following dispersion
equation providing nontrivial solutions for $\varepsilon_y$:
\begin{equation}\label{Q}
\frac{ 1-\lambda-\tau (k_y^2+k_x^2)}{n\lambda+n
\tau (k_y^2+k_x^2)+ 1} = \frac{k_x^2}{k_y^2+n\lambda} \, .
\end{equation}
The corresponding quadratic equation for $\lambda (k_x,k_y)$ has
the form
\begin{equation}\label{SqEq}
\lambda^2+P\lambda+Q=0\, ,
\end{equation}
where
\[
P=k_x^2 + k_y^2/n - 1 + \tau (k_y^2+k_x^2)\, ,
\]
\vspace*{-4mm}
\begin{equation}\label{Deno1}
Q=\frac{k_x^2-k_y^2}{n} + \tau \left(k_x^4 + \frac{n+1}{n}\, k_x^2
k_y^2 + \frac{1}{n}k_y^4 \right)\, .
\end{equation}
The system is unstable if $\Re \lambda(k_x,k_y)>0$.

\subsection{Qualitative Analysis}

The dispersion equation becomes more transparent when the heat
conductivity can be neglected, i.~e. $\tau=0$. Then,
\begin{equation}\label{tau0}
\lambda^2+\lambda (k_x^2 + k_y^2/n - 1) + (k_x^2-k_y^2)/ n =0\, .
\end{equation}
First, we notice that at $k_x=0$ the system is
always unstable. This is not surprising since such solutions
correspond to the case of a sample with fixed transport current,
$ i_y=0$, heated by the electric field $E$ under
adiabatic conditions. In this case $\delta E$ and $\delta T$ grow
with the maximal possible rate, $\lambda=1$, and the
characteristic time of the instability build-up is $t_0$.
\begin{figure}[ht]
\includegraphics[height=5.2cm]{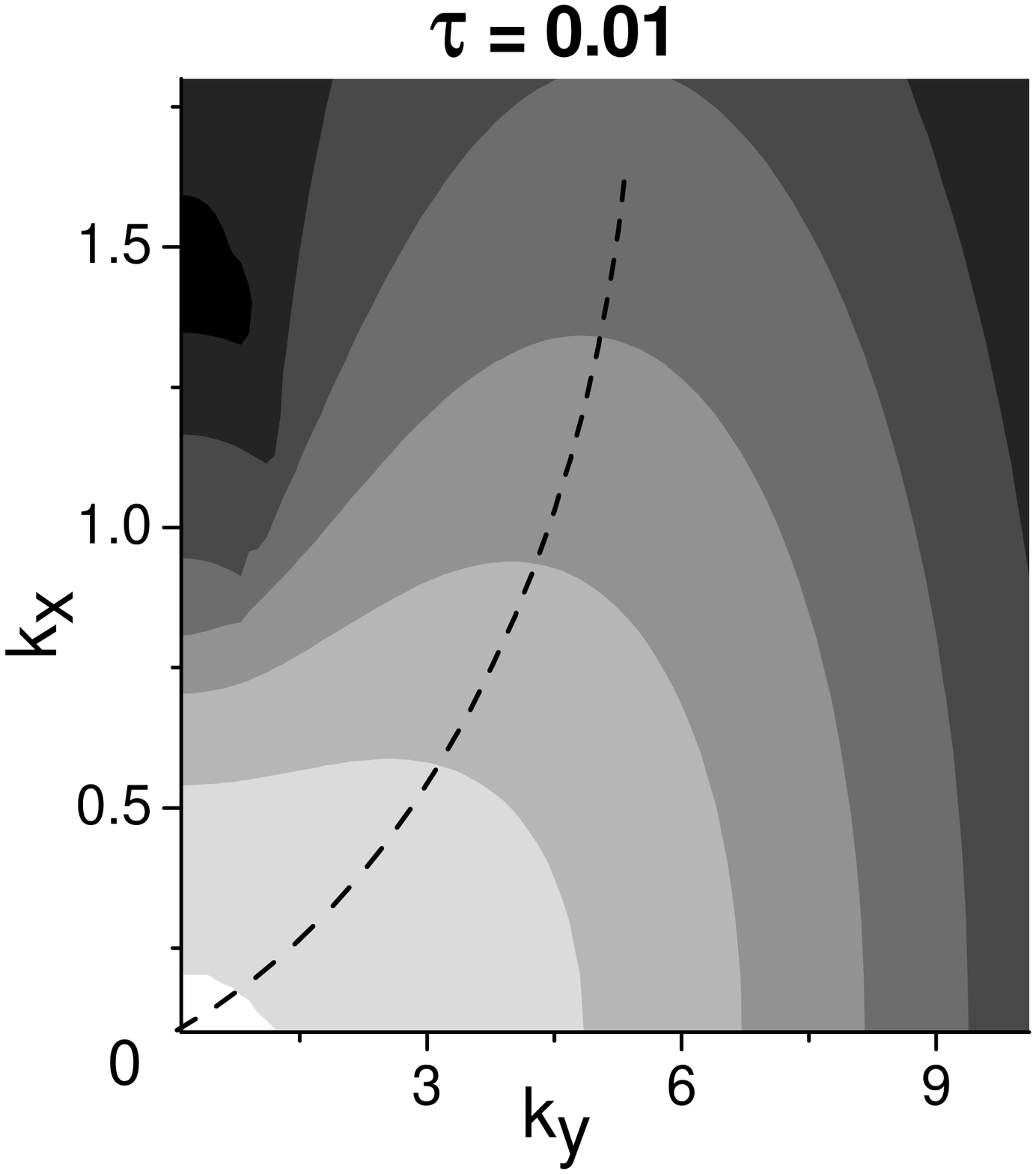}
\includegraphics[height=5.2cm]{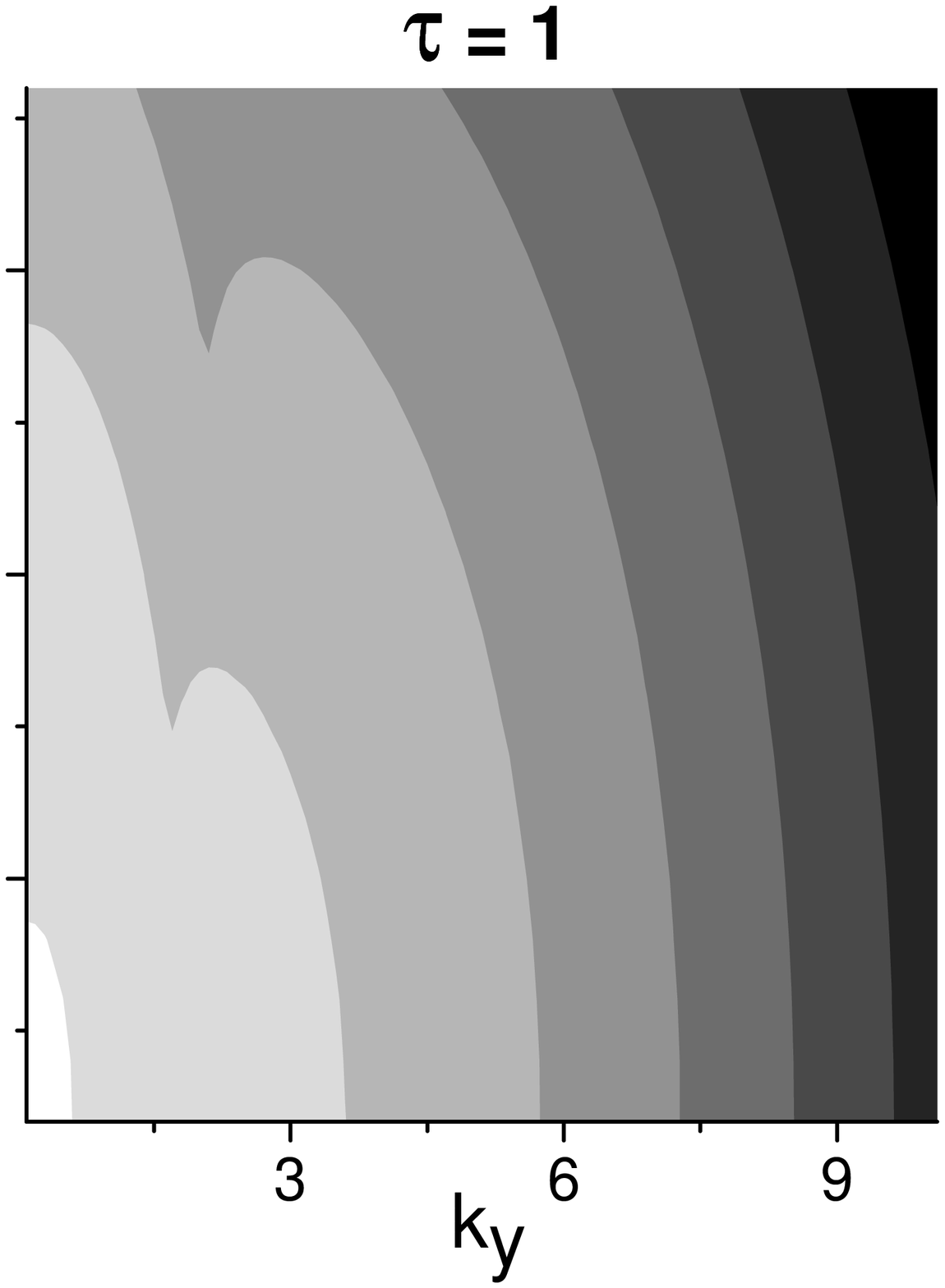}
\caption{\label{fig23} The contour plots for the instability
increment $\Re\lambda(k_x,k_y)$ obtained from \eq{SqEq} for
$n=10$. The brightest areas correspond to the fastest growth of
instability. For low $\tau$ perturbations with a finite $k_y$ have
the maximal increment, while for large $\tau$ (strong heat
diffusion), uniform perturbations with $k_y=0$ would grow
fastest.}
\end{figure}

For a finite sample the $k_x$ is not arbitrary
because of the boundary conditions at the edges of the flux
penetrated region. Only some particular $k_x$ satisfy the boundary
conditions, which makes the system more stable. For example, for
perturbations uniform in the $y$-direction ($k_y=0$) the
instability develops only if $k_x<1$. However, if we set $k_y
\rightarrow \infty$, then the system becomes unstable for any
$k_x$, and we again arrive at the maximal growth rate,
$\lambda=1$. This result can be understood physically, if we take
into account that infinite $k_y$ correspond to a perturbation in
the form of a narrow finger directed along the $y$ axis, i.~e.,
perpendicular to the current flow. The current flow remains
unperturbed by an infinitesimally narrow finger i. e.,
the condition $i_y=0$
lest favorable for the stability holds, like for the case $k_x=0$.
In the case of wider fingers, the current adjusts itself to
the temperature fluctuation, which slows down the instability
growth. So, if one neglects the thermal diffusion, the narrowest
possible fingers are the most favorable 
($k_y \rightarrow \infty$),
and the superconducting state is utterly unstable.

The thermal diffusion evidently suppresses the instability growth.
The suppression is most effective for large $k_y$. As a result, we
obtain some optimal value of $k_y$, for which the instability
increment $\lambda$ is maximal. The existence of such an optimal
$k_y$ is evident from the contour plot of $\Re \lambda$ calculated
for $\tau=0.01$, see \f{fig23}, left. The dashed line shows $k_y$
providing the maximal $\Re \lambda$ for a given $k_x$. However, if
$\tau$ is larger then the heat diffusion fully dominates the
instability development. In that case the maximal $\lambda$
corresponds to $k_y=0$, see \f{fig23}, right.

\section{Results}

In this section we solve the problem more accurately, and  start
by establishing the proper boundary conditions.

\subsection{Boundary conditions}

From the above analysis it is clear that a finger structure may
appear only for  $\tau < 1$. Consequently, we focus only on this
case. Since the thermal diffusion is then slower than the magnetic
diffusion, we can impose only the electrodynamic boundary
conditions. This is equivalent to neglecting the heat flux in the
$x$ direction, i.~e., the term $\tau k_x^2$ in \eq{thL} can be
omitted.

The magnetic field at the slab surface is equal to the applied
field, hence the perturbation at the surface is zero,
$\delta h_z = 0$ at $x=0$.
The magnetic field has only $z$-component, thus from Eqs.(\ref{E})
and
(\ref{Ex}) one obtains $\delta E'_y \propto \delta h_z$, and
the first boundary condition is
\begin{equation}\label{B1}
\delta E'_y = 0,\,\,\,x=0 \, .
\end{equation}
This condition also means that the current does not flow across
the sample surface, $\delta j_x \propto \delta E_x=0$ at $x=0$.

Let us specify the boundary conditions at the flux front, $x=l$.
In the flux-free region, $x>l$, the electric field  decays on the
scale of the London penetration depth, which is much
smaller than any spatial scale of the problem. Therefore, the
continuity of the tangential component of the electric
field requires,
\begin{equation}\label{B2}
\delta E_y = 0,\,\,\,x=l \, .
\end{equation}
These boundary conditions together with \eqss{i}{thL}{Eldyn} are
satisfied when $\delta E_y \propto \cos (k_x x/w)$ with
$$k_x=(\pi/2)\, (w/l)\, . $$
Now we can search for solutions of
\eq{SqEq} with this $k_x$, and as before, when $\Re \lambda >0$
the system is unstable.

\subsection{Instability criterion and increment}

Let first consider the spatially uniform
case where there exists a well-known criterion for the
thermomagnetic stability.\cite{MR,GMR,wipf91,chapter,swartz,Wipf}
With $k_y=0$ and
using $\tau \ll 1$, i. e. for very slow thermal diffusion
we find from \eq{SqEq} that the system is unstable if
$k_x < 1$. For the Bean model, where $ l =H/j_c $,
this is expressed as
\begin{equation}\label{ad}
H > H_{\text {adiab}} =(\pi/2) \sqrt{CT^{*}/\mu_0}\, ,
\end{equation}
which is the commonly used adiabatic criterion
for flux jumps.

\begin{figure}[ht]
\includegraphics [width=8.0cm]{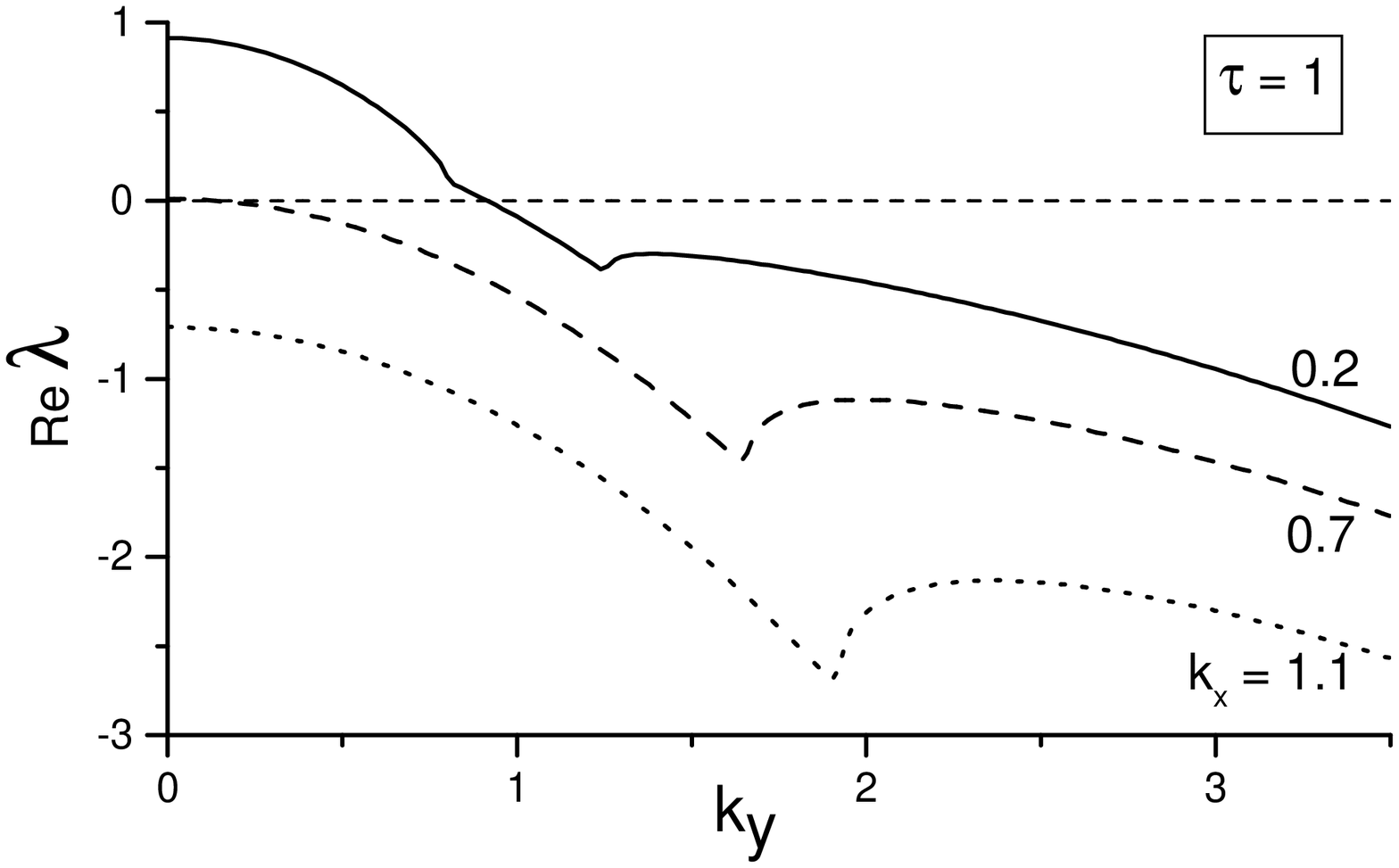}
\includegraphics [width=8.0cm]{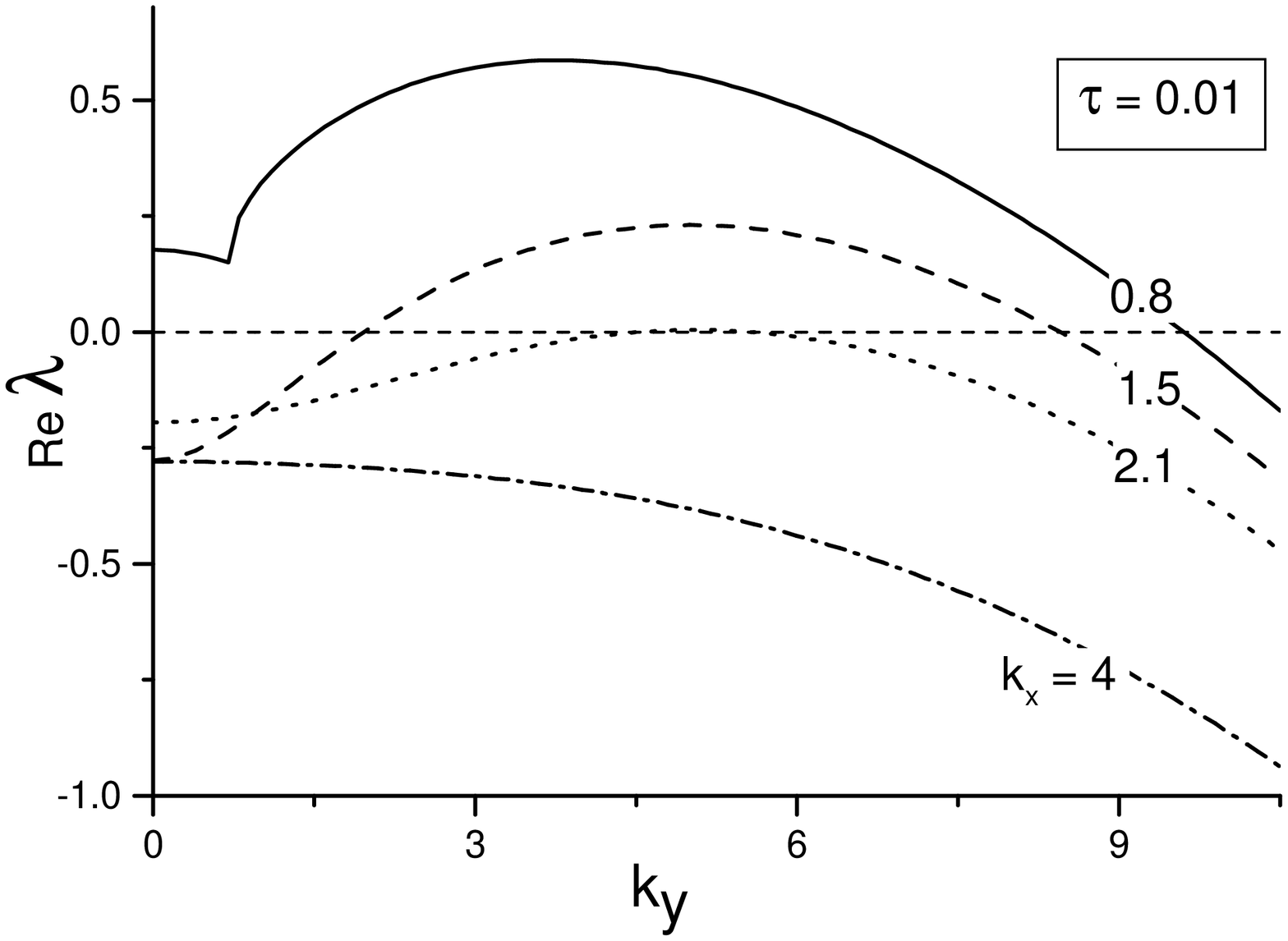}
\caption{\label{f_ky} The instability increment $\Re\lambda(k_y)$
found from \eq{SqEq} for $n=10$ and different $k_x$. Top: fast
heat diffusion, the
maximal increment corresponds to uniform perturbations ($k_y=0$).
Bottom: slow heat diffusion, the maximal $\Re \lambda$ is found at
a finite $k_y$. }
\end{figure}
Let us next consider cases of non-zero $k_y$, and analyze the
behavior of $\Re \lambda(k_y)$. Shown in \f{f_ky} (top) are plots
for $\tau=1$. For small applied magnetic fields the system is
stable, see the curve for $k_x=1.1$. As the field increases, the
flux penetration depth grows, and hence $k_x$ goes down. For
$k_x=0.7$ the system becomes unstable, i.~e., solutions with $\Re
\lambda>0$ arise. Note that the instability appears first at
$k_y=0$. For higher fields ($k_x=0.2$), the instability range
extends to large $k_y$ too, but the maximal $\Re \lambda$ always
corresponds to $k_y=0$.\cite{kink} Therefore, for relatively large
$\tau $ the instability develops in a uniform mode.

However, for smaller $\tau$ the $\Re \lambda(k_y)$ behaves
differently, see \f{f_ky} (bottom). The maximal $\Re \lambda$ can
here occur for a non-zero $k_y$. Moreover, it is possible that the
system is stable with respect to uniform perturbations, while
unstable for perturbations with finite $k_y$, see the curve for
$k_x=1.5$ . This means that a non-uniform structure along the
$y$-direction will be formed.

When the applied magnetic field gradually increases from zero, the
instability first starts for some particular $k_x=k_x^*$ when $\Re
\lambda=0$ only for one single value of $k_y=k_y^*$. This is the
case for $k_x=k_x^*=2.1$ in \f{f_ky} (bottom).
To find these $k_y^*$ and $k_x^*$ one needs two conditions.
The first one is
\be
Q(k_x^*,k_y^*)=0,
\ee
which is a quadratic equation
with respect to $(k_y^{*})^2$, and the second one is that the
discriminant of this equation is zero.
Using \eq{Deno1} and the fact that $n \gg 1$, we find
\be
 k_y^* = \l(\frac{2}{n} \r)^{1/4} \frac{1}{\sqrt\tau}\,
 , \quad
  k_x^* = \frac{1}{\sqrt{n\tau}}  \, ,
\label{kyopt}
\ee

The instability occurs at $k_x < k_x^*$,
and for large $n$ this instability criterion can be written
as
\be
 E > (\pi^2/4) \, (\kappa T^*/j_c l^2)\, .
 \label{ins}
\ee
One can see from \f{f_ky} (bottom) that the value of $k_y$
where $\Re \lambda$ has the maximum depends only weakly on $k_x$.
Therefore, a good estimate for the finger width $d_y$ in the $y$
direction is $w/k_y^*$. Thus
\be
  d_y \approx
  \l( \frac{\kappa}{E\ \partial j_c/ \partial T}\r)^{1/2} \! \!
\frac{1}{(2n)^{1/4}}
  \, .
\ee
Once we go from the instability threshold towards lower $k_x$,
the increment $\Re \lambda$ quickly becomes of the order of unity.
Thus, the characteristic time of the instability development is of
the order of the adiabatic time, $t_0$.

The aspect ratio of the perturbed region is
 \be
  k_y^*/k_x^*
\approx (2n)^{1/4} \, .
\ee
Note that it is independent of the thermal parameters,
$C, \kappa, T^*$, and determined only by the shape of the $E(j)$
curve.

As was seen from \f{f_ky}, the instability will
develop uniformly for $\tau=1$,
and non-uniformly for $\tau=0.01$.
It follows directly that the border between the uniform and
non-uniform regimes is given by the criterion
$\Re \lambda(k_x^*,k_y=0)=0$.
Using \eqs{SqEq}{kyopt} one can rewrite the criterion as 
$\tau = 1/n$.
Rewriting this in dimensional form we conclude that for
\be
E > E_c = \mu_0 \kappa j_c/C \,  \label{domain}
 \ee
the instability will evolve non-uniformly.
\section{Simulations}
\begin{figure}[ht]
\includegraphics[width=8.5cm]{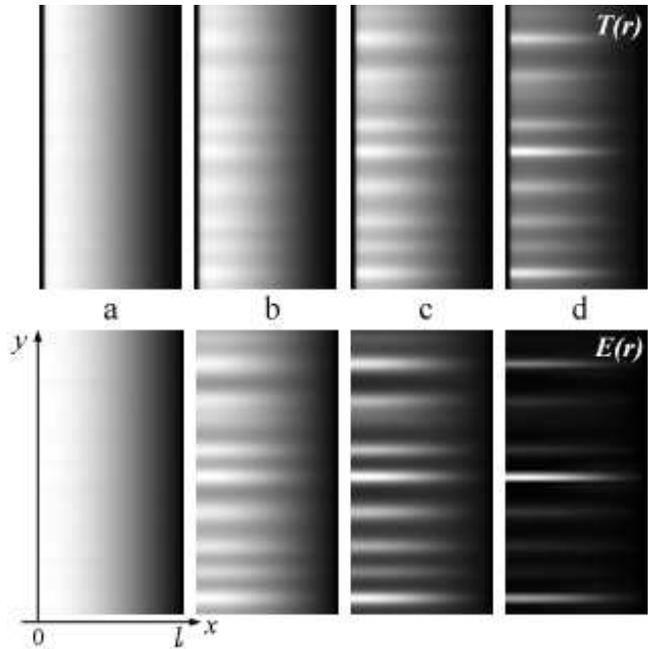}
\caption{\label{f_e} Evolution of the temperature $T$ and electric
field $E_y$ distributions produced by simulations that illustrate
the formation of a finger structure.
The instability was triggered by a uniform  electric field,
$E_y=E_0$, switched on at $t=0$. The images (a--d) correspond to
the  times $t/t_0=1.6, 3.0, 3.2$, and $3.3$, respectively.}
\end{figure}
\begin{figure}[ht]
\includegraphics[width=8.5cm]{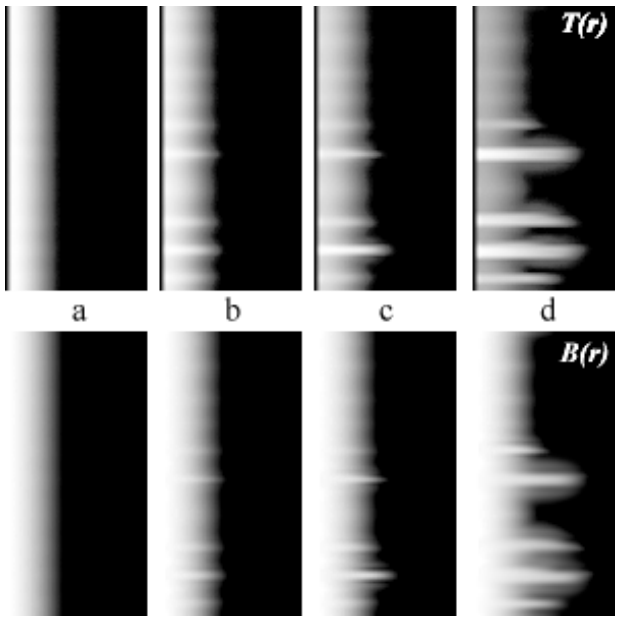}
\caption{\label{f_tb} Evolution of the temperature and flux
density distributions produced by simulations.
The instability is triggered by applying a magnetic field. The
images (a--d) correspond to the times $t/t_0=45, 46.1, 46.15$, and
$46.25$, respectively. The formation of fingers and their
propagation into the flux free area is clearly seen.}
\end{figure}
In order to visually illustrate the formation of non-uniform
structures, and to verify the validity of the above analytical
results, numerical simulations based on the Maxwell and thermal
diffusion equations \eqss{B}{th}{E} were carried out.  In the
simulations we went beyond the linear approximation and considered
the full non-linear $E(j)$ curve, which was chosen to be
\be
E = \frac{j}{\sigma_{f} + (j_c/j)^{\tilde{n}-1}\ j_{c0}/E_0}\,
, \label{fullCVC}
\ee
where $j_{c0}$ and $E_0$ are constants. This
is one of possible smooth interpolations between the flux creep
regime at small currents with $E \propto j^{\tilde{n}}$, and the
Ohmic flux flow regime $E = j/\sigma_{f}$ at high $j$. Here the
flux flow resistivity is much higher than the characteristic
resistivity in the flux creep regime, $\sigma_{f}^{-1} = 10^4
E_0/j_{c0}$. The temperature dependence of the critical current
density is assumed to be linear, $j_c = j_{c0} [1 - (T-T_0)/T^*]$.
The
electrodynamic boundary conditions are $dE_y/dx(x=0)=0$ (constant
external magnetic field), $E_y(x=l)=0$, $E_x(x=0,l)=0$, with
$l=2w$, and the periodic boundary conditions in the $y$ direction.
Since the thermal boundary conditions are not of crucial
importance at $\tau\ll 1$, we used the simplest ones,
$T(x=0)=T_0$ (ideal heat removal at the surface) and
$dT/dx(x=l)=0$ (the symmetry condition in the middle of the slab).

Our analytical results predict that the instability will form a
finger structure if a sufficiently large uniform background
electric field is present. Therefore, for initial conditions we
assume a uniform electric field, $E_y(0\le x <l,t=0)=E_0$.
To introduce some non-uniformity into the system small
random values $\delta T_R \ll T^*$
were added to the initial temperature
for every discrete
node.
The initial temperature is then given as 
$T(x,t=0) = T_0+\delta T_R$.
The key parameters $\tau=0.001$ and $n \approx \tilde{n}=30$
are specified at $E=E_0$. Their dependences on the electrical
field are given by \eqss{n}{sigma}{fullCVC}. Since now 
$E_0 = E_c/(n \tau) \gg E_c$, 
the condition (\ref{domain}) is fulfilled,
and the instability is expected to develop in a non-uniform
fashion. This is indeed confirmed by the calculated evolution of
$E_y$ and $T$ distributions presented in \f{f_e}.
The numerical solution was performed on a grid of
140$\times$70 nodes using a simple-step integration method.

One can see from \f{f_e}(a) that at small times
the distributions of $E$ and $T$ are essentially uniform
along the $y$ axis.
Then, a finger structure is emerging (b) with protrusions
perpendicular to the electric field direction, as predicted by our
previous linear analysis. The simulations also show how this
finger structure is evolving beyond the linear regime. We can see
that the electrical field in some fingers grows faster so that
relative difference between the fingers increases
 (c). Eventually, the most intense finger takes over and
dominates the entire $E$ distribution (d). We believe that the
reason for such behavior is the increase of the differential
resistivity as $E$ grows, \eq{fullCVC}. The growth of $E$ is
significant: the average value $\bar{E_y}=1.7E_0$ for (a), and
$21E_0$ for (d). Note that this growth cannot be traced from the
presented images only because the gray scale was  optimized for
each individual image to provide the best contrast. More detailed
simulations showed that the instability growth slows down only
when the increasing $E$ reaches the inflection point on the $E(j)$
curve before entering the flux flow regime.

Next, we carry out simulations with different initial and boundary
conditions. We start from zero electric field, $E(t=0)=0$, and
assume that a linearly increasing magnetic field is applied to the
slab so that $-dE_y/dx=dH/dt=0.03wj_c/t_0$ at $x=0$. The other
parameters are the same except that now the slab halfwidth is
$6w$.
The right edge of the distributions shown in \f{f_tb}
corresponds to the middle of the slab.
One can see from \f{f_tb}(a) that for
small $H$ the flux penetrates in the conventional way, and a
Bean-like profile of flux density is gradually advancing into the
slab. When $H$ and correspondingly $E$ increase further, an
instability sets in and leads to the formation of fingers (b-d).
The finger structure is apparent  in both the $B$ and $T$
distributions, especially on the later stages when only few
intense fingers remain. Remarkably, the fingers tend to propagate
into the flux free region, strongly distorting the flux front (d).
One can speculate that  these growing fingers eventually will
develop into a complex dendritic flux pattern observed by
magneto-optical
imaging.\cite{1967,leiderer,bolz,bolz03,duran,vv,epl,sust,apl,prb,phc04,rudnev}

The instability criteria and its growth rate found from the
simulations are in a good agreement with our analytical results.
Moreover, the simulations demonstrate that the finger instability
arises even if some assumptions made in the derivation are
relaxed. In particular, one does not necessarily need a strictly
uniform background $E$ and $T$ distributions as assumed in the
derivation. Furthermore, the background $E$ and $T$ distributions
can also be non-stationary, which is always the case in a real
experiment. In fact, in the simulations relevant to \f{f_tb},
where the instability was triggered by increasing the applied
magnetic field, the $E$ and $T$ distributions were non-uniform and
non-stationary. The formation of finger structure also turned out
to be rather insensitive to the boundary conditions. We have also
carried out simulations assuming that $j_c$ in \eq{fullCVC}
depends on the local $B$ according to the Kim model,\cite{kim}
$j_c(B) \propto (B_0+|B|)^{-1}$. With $B_0=3\mu_0wj_c(0)$ we
found similar distributions, thus proving  that the finger
instability can arise also in cases with a $B$ dependent $E(j)$.

The simulations presented here have some similarities with those
by Aranson \textit{et al}.\cite{GV} The main differences are that
Aranson \textit{et al.} started from a fully-penetrated state, the
instability was nucleated by a local heat pulse, and $j_c$ was
generally nonuniform. As a result, the obtained patterns of $T$
distribution look different from ours. Nevertheless, they also
found that the instability results in a non-uniform $T$
distribution only at small $\tau$.

\section{Discussion} \label{Disc}

\begin{figure}
\includegraphics [width=8.5cm]{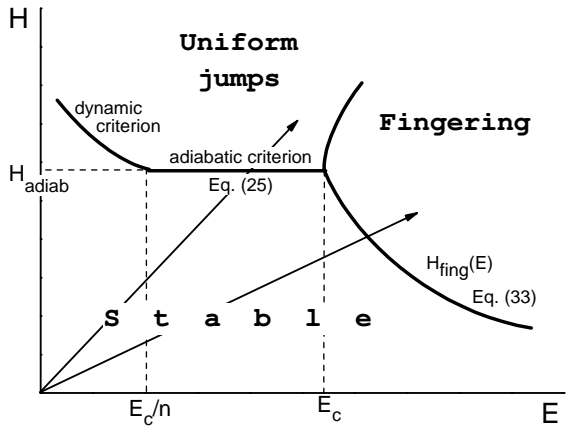}
\caption{\label{f_he} Instability phase diagram in the plane
magnetic field -- electric field. The horizontal line corresponds
to the adiabatic criterion for uniform jumps, \eq{ad}. For
$E>E_c$, the instability has a finger structure, and the criterion
is given by \eq{Hd}.}
\end{figure}

The results obtained in this work can be graphically summarized by
the instability ``phase diagram'' shown in \f{f_he}. For small
electric fields, $E<E_c$, the conventional uniform instability is
favorable, and the adiabatic instability criterion \eq{ad} is
applicable. For $E>E_c$ the fingering instability develops, with
the instability criterion given by \eq{ins}. Using the Bean model,
$H=j_c l$, we obtain the finger instability criterion as
\begin{equation}\label{Hd}
    H > H_{\text{fing}} =
    (\pi/2)\sqrt{\kappa T^* j_c/ E}\, ,
    \quad E > E_c \, .
\end{equation}
Figure also shows the border between the regions of uniform jumps
and
fingering instability for $H>H_{\text{adiab}}$
that was calculated
from \eq{SqEq}
using two conditions,
$\partial \Re \lambda/\partial k_y\vert_{k_y=k_0}$=0, and
$\Re \lambda(k_x,k_0)=\Re \lambda(k_x,0)$.

Strictly speaking our analysis applies to the case $\tau < 1$,
which is equivalent to $E>E_c/n$. For smaller electric fields a
similar stability analysis can be made, taking into account the
heat flux along the $x$ axis. As expected, we found that for
$E<E_c/n$ the uniform development of instability is always
preferable. The instability criterion is given by the well-known
dynamic criterion, that is highly sensitive to the external
cooling conditions.\cite{MR,GMR} However, in all cases the flux
jump field decreases monotonously with $E$, as indicated
schematically in \f{f_he}.

The finger instability occurs only at rather large background
electric field. This field can be created by different sources.
For example, if the applied magnetic field increases with a rate
$\dot{H}$, an electric field $E \sim \dot{H} l \sim \dot{H} H
/j_c$ is generated. Thus, when increasing $H$ with a constant rate
we move in the phase diagram in \f{f_he} along a straight line
staring from the origin. For small $\dot{H}$, one crosses the
instability boundary at $E<E_c$, resulting in a uniform flux jump.
For large $\dot{H}$, the stability is destroyed for smaller $H$,
and results in the formation of a non-uniform spatial structure.
The predicted downturn of the  $H(E)$ instability line at large
$E$
can be checked experimentally.

Numerical estimates were made using typical parameters for
low-temperature superconductors at helium temperatures: $j_c =
10^{10}$ A/m$^2$, $C = 10^3$ J/Km$^3$, $T^* = 10$~K, $\kappa =
10^{-2}$ W/Km, and $n  = 30$. We then find the following values
for the characteristic fields, $H_{\text{adiab}} \approx 0.1$~T,
and $E_c \approx 0.1$~V/m, a finger width of $d_y \approx 3\,
\mu$m for $E\sim E_c$, and a build-up time of the instability,
$t_0$, in the $\mu$s range. These estimates are not far from those
reported in experimental papers, namely dendritic fingers of
width~20-50~$\mu$m,\cite{leiderer,duran,sust,prb} and the
instability build-up time of 
$\sim 0.1\,\mu$s.\cite{leiderer,bolz,bolz03}
The criterion for the fingering instability $E>E_c$ can also be
written down as
$\sigma < \sigma_c = C/n \mu_0 \kappa$.
Using the numbers above we find
$\sigma_c = 3 \times 10^9 \ \Omega^{-1}m^{-1}$, which is a
reasonable value
for the flux-flow conductivity.
Correspondingly, for 
$\sigma=\sigma_c$ one obtains $\tau=1/n \sim 1/30$.

Note that the electric field, $E_c$, needed for the finger
instability to occur is not very small. As an estimate, the
magnetic field ramp rate $\mu_0\dot{H}$ that induces the electric
field $E_c$ is of the order of 10$^2$~T/s for $l$=1~mm. Rates of
similar magnitude are conventionally used for pulsed magnetization
of  superconducting permanent magnets.\cite{sander02,surzhenko}
In experiments reporting the fingering
instability\cite{1967,leiderer,bolz,bolz03,duran,vv,epl,sust,apl,prb,phc04,rudnev}
the ramp rates were much smaller.
One should keep in mind however that
the actual electric field can be much larger than it follows from
the
simple estimate $\mu_0\dot{H}l$. The reason is a strong
non-uniformity of the flux penetration both  in space and in time,
see for review
Ref.~\onlinecite{aval}. Hence, one can
expect rather large local electric fields that last longer than
the inverse instability increment, $\gtrsim 1 \ \mu$s.
Other sources of large electric fields include random fluctuations
of the superconductor parameters due to, e.~g.,
relaxation of mechanical stresses.
A very large electric field can  be also created on
 purpose, e.~g., by a laser pulse, which nucleates highly
nonuniform flux
distributions.\cite{leiderer,bolz,bolz03}

In any case, it is rather difficult
to meet the fingering instability criterion, $E > E_c$. This might
be
the reason why fingering is hardly observed in bulk samples.
We are aware of only one experimental work\cite{1967} where
an indication of the discussed fingering instability in
relatively thick samples (with thickness up to 2~mm)
was obtained. Another possible reason for why such observations
are few is that flux jumps in bulk superconductors are often
complete, or almost complete. This means that the temperature
rises close to $T_c$ in the entire sample, leading to a uniform
flux distribution which erases any trace of a possible
non-uniformity in the first stages of the flux jump. That
contrasts the behavior in thin film samples, where the jumps are
usually much smaller and far from being complete.\cite{zhao} This
makes non-uniform  jumps easier observable in
films.\cite{leiderer,bolz,bolz03,duran,vv,epl,sust,apl,prb,phc04,rudnev}
Moreover, huge
stresses usually exist between a superconducting film and
substrate. Abrupt relaxation of these stresses can lead to
fluctuations in $E$, especially for Nb$_3$Sn or MgB$_2$ where the
superconducting properties depend strongly on the
strain.\cite{GMR,mgb2review}
Note also that in films it is much more probable that any
perturbation
of electric field will influence the whole thickness, whereas
it will affect only a small part of a bulk sample.
Although our equations for a slab cannot
be directly applied to the case of a thin film, we expect that
essentially the same physics describes the formation of finger
structures in the films too.\cite{aranson04} Moreover, non-
locality of the
current-field relations in films can make formation of non-uniform
structures there even more favorable, and possibly account for the
branching flux patterns observed experimentally.

The presence of a background electric field $\mathbf{E}$, and
hence moving magnetic flux implies that the background state
itself is not stationary.
In a typical experiment, the applied magnetic field is
increasing $H=H(t)$, the flux front is moving into the sample,
$l=l(t)$,
and hence the electric field is non-stationary within the
flux-penetrated region.
Obviously, our analytical results are
valid only if all these quantities
change in time slower than the
perturbations $\delta \mathbf{E}, \delta T$ grow, i.~e.,
when $\dot{E}/E,\ \dot{H}/H,\ \dot{l}/l \ll \lambda$.
If the electric field is created by ramping the external
magnetic field,   
$\mathbf{E}(x,t) \simeq \mu_0 {\dot H} [l(t)-x]$, then
$\dot{E}/E \approx \dot{H}/H\approx \dot{l}/l$. 
Using that $\lambda \sim 1/t_0$, we can
rewrite the above inequality as 
$H \gg H_{\text{adiab}}/\sqrt n$. Since $n \gg 1$ this condition is
satisfied in the major part of the phase diagram in \f{f_he}.

In conclusion, a linear analysis of heat diffusion and Maxwell
equations shows that a thermomagnetic instability may result in
finger-like distributions of $T$, $E$ and $B$. The fingering
instability arises if the background electric field is so high
that the magnetic flux diffusion proceeds much faster than the
heat diffusion. Numerical simulations have shown that upon further
development of the instability one finger starts growing much
faster than the others, and propagates into the flux-free region.

\begin{acknowledgments}

This work is supported by INTAS grant 01--2282, NATO Collaborative
Linkage Grant 980307 and FUNMAT\verb+@+UiO. We are thankful for
helpful discussions with B.~Shapiro, V.~Vinokur, P.~Leiderer and
L. M. Fisher. The work in Houston is supported in part by the State of Texas through the
Texas Center for Superconductivity and Advanced Materials at the University
of Houston.

\end{acknowledgments}


\begin{thebibliography}{99}
\bibitem{kink} Most curves in \f{f_ky} show a kink when
the discriminant of \eq{SqEq} changes sign, which marks appearance
of a non-zero $\mathop{\mathrm{Im}} \lambda$.
\end{thebibliography}
\end{document}